\documentclass[12pt,preprint]{aastex}
\shorttitle{Spectroscopic Investigation of Crab Nebula}
\shortauthors{MacAlpine et al.} 
\received{2006 July 2}
\begin{document}

\title{A SPECTROSCOPIC STUDY OF NUCLEAR PROCESSING\\
AND PRODUCTION OF ANOMALOUSLY STRONG LINES\\ IN THE CRAB NEBULA
\footnote{This paper involves data obtained at the 
Michigan-Dartmouth-MIT Observatory and at the McDonald Observatory 
of The University of Texas at Austin.}
}

\author{Gordon M. MacAlpine, Tait C. Ecklund, William R. Lester, Steven J. Vanderveer}
\affil{Department of Physics and Astronomy, Trinity University, San Antonio, TX 78212}

\and

\author{Louis-Gregory Strolger}
\affil{Department of Physics and Astronomy, Western Kentucky University, Bowling Green, KY 42101}

\begin{abstract}
We present and discuss correlations for optical and near-infrared ($5500-10030$~\AA)
line intensity measurements at many positions in the Crab Nebula.
These correlations suggest the existence of gas produced by a range 
of nuclear processing, from material in which synthesis ended 
with the CNO-cycle, to some helium-burning and nitrogen depletion, 
to regions containing enriched products of oxygen-burning.  
The latter exhibit a gradual, linear rise of $[$Ni~II$]$ emission with increasing argon enrichment, 
whereas gas with less nuclear processing shows markedly different  
$[$Ni~II$]$ emission characteristics, including the 
highest derived abundances.  This suggests two origins for 
stable, neutron-rich nickel in the nebula: a type of ``alpha-rich freezeout''
in the more highly processed material, and possibly removal of ions from the neutron star
in other regions.  In addition, the data indicate 
that anomalously strong observed $[$C~I$]$ emission comes from broad, low-ionization 
H$^{+}$ to H$^{0}$ transition zones.  Although the strongest 
He~I emission could also be enhanced in similar low-ionization gas, 
correlations between relevant line ratios argue against that interpretation, strengthening the case 
for an exceptionally high helium mass fraction in some locations.
\end{abstract}


\section{INTRODUCTION}

Young supernova remnants are excellent laboratories for investigating 
how stars make elements, and the Crab Nebula in particular can provide unique 
information about the precursor star, the supernova event, associated heavy element production, 
and the environment of a highly energetic pulsar.  It is the bright remnant 
of a core-collapse supernova observed in 1054~A.D.  Its age and location, 
roughly 180~pc away from the plane of the Galaxy, suggest that the ejecta 
are not significantly contaminated by swept-up interstellar material.  
Furthermore, measured electron temperatures in the gas  (Woltjer 1958;
Miller 1978; Fesen \& Kirshner 1982; MacAlpine et al. 1989, 1996) along with 
the lack of other possible evidence for shocks (Frail et al. 1995) imply that the line-emitting 
gas shines primarily because of photoionization by the locally-generated 
synchrotron radiation field (see also Davidson \& Fesen 1985).  Therefore it can be analyzed using powerful 
numerical photoionization modeling codes.  It is generally believed that the 
supernova precursor star initially contained about $9-11$~M$_{\odot}$ (e.g.,
Arnett 1975; Nomoto 1985), representing the important low end of the Type~II 
supernova mass range, below which stars could ignite carbon degenerately and 
above which successive nuclear reaction stages would be expected to take place 
through silicon burning.  As discussed by Woosley \& Weaver (1986a), the 
applicable stellar models allow for a number of explosive and nucleosynthesis 
possibilities.

Spectroscopic and photometric investigations of the Crab Nebula to date have 
indicated several apparent ``gas components.''  The majority of the observed 
nebular gas is helium (e.g., MacAlpine et al. 1989), consisting of less than 
$2$ M$_{\odot}$ (MacAlpine \& Uomoto 1991).  It has been postulated that this 
represents ``helium mantle'' material from deep within the original star, 
ejected by the explosive event.  This is consistent with some stellar models, 
or scenarios involving another star, in which outer layers of the precursor were 
lost prior to the core-collapse event.  Most of this helium-mantle gas 
appears to be nitrogen-rich, confirming its origin from CNO processing 
(MacAlpine et al. 1996).

There also is a major component of gas, primarily in the southern part of the nebula
near the pulsar, which is significantly nitrogen-poor and much of which is sulfur-rich 
(MacAlpine et al. 1996).  It was suggested that this gas resulted from localized 
oxygen-burning episodes, consistent with stellar models (Woosley \& Weaver 1986b, 1995)
that involve off-center (in a shell) oxygen-burning.  Strolger \& MacAlpine (1996)
provided a preliminary demonstration that this explanation is probably correct.

Still another significant nebular component or ``anomaly'' involves an apparent 
helium-rich torus viewed as an east-west band across the pulsar region, which constitutes 
approximately 25\% of the visible material (Uomoto \& MacAlpine 1987; MacAlpine et al.
1989; MacAlpine \& Uomoto 1991).  
The computed helium mass fraction is about 95\%, 
and there is not yet a realistic explanation for this apparent structure.

Other known apparent anomalies in the Crab Nebula include exceptionally strong $[$Ni~II$]$
and $[$C~I$]$ line emission, resulting in large spatial variations for deduced 
abundances of nickel (along with iron) and carbon.  Strong $[$Ni~ II$]$~$\lambda$7378 has 
been reported by numerous authors (Miller 1978; Fesen \& Kirshner 1982;
Henry et al. 1984; MacAlpine et al. 1989).  
The latter work suggested neutron-rich 
nickel isotopic abundance enhancements, compared with solar, by factors of 5-50 at 
various locations.  Iron was also found to vary widely, together with nickel 
but at a much lower level; deduced nickel/iron abundance ratios are roughly 60-75 
times the solar value.

Henry et al. (1984) found surprisingly high $[$C~I$]$~$\lambda$$\lambda$9823,9850 emission at several 
locations in the Crab Nebula, where it was measured to be as much as 7 times stronger 
than predicted by the photoionization models of Henry \& MacAlpine (1982).  To 
account for this, it was postulated that the $[$C~I$]$ lines might arise from 
collisional excitation involving hydrogen atoms as well as electrons.

As part of an effort to develop a more consistent and accurate overall understanding
of the Crab Nebula, we have measured relative line intensities of He~I~$\lambda$5876,
$[$O~I$]$~$\lambda$6300, H$\alpha$, $[$N~II$]$~$\lambda$6583, $[$Ar~III$]$~$\lambda$7136, 
$[$Ni~II$]$~$\lambda$7378, $[$S~III$]$~$\lambda$9069, $[$S~III$]$~$\lambda$9531, 
and $[$C~I$]$~$\lambda$9850 for roughly 200 
well-distributed positions throughout the emitting gas.  In this paper, we address 
all of the above issues (apparent gas components and anomalies) using comparisons among 
these line measurements.  A follow-up paper, involving more in-depth photoionization 
analyses for deriving improved gas physical conditions and chemical abundances, 
is planned.

The spectroscopic observations are described in \S~2. Then \S~3 presents 
emission-line correlations and resulting inferences about the nebular gas.
The correlations illustrate a broad range of nuclear processing and confirm
significant enrichment with products of oxygen-burning in some areas. 
They also suggest two distinct nickel/argon line relationships. 
In addition, the importance of optical depth for enhanced $[$C~I$]$ lines is demonstrated, 
along with the {\it lack} of a dominant role for optical depth in the production of strong 
He~I emission.  A summary discussion is given in \S~4.

\section{OBSERVATIONS}

Spectroscopy covering the wavelength range from approximately 5500 to 7700~\AA\ was 
obtained through a long slit at various orientations across the pulsar 
during the nights of 1995 January 21 and 22, at the 2.4-m Hiltner 
telescope of the former Michigan-Dartmouth-MIT Observatory, 
which is located on Kitt Peak.  The Mark~III spectrograph was 
used with a TEK~1024$\times$1024 CCD and a 600~lines~mm$^{-1}$ grism blazed 
at 5800~\AA, resulting in roughly 2.3~\AA~pixel$^{-1}$ dispersion.  
The slit width was 1\farcs2, projected on the sky, and the effective 
projected length was about 4\farcm5.  The slit positions employed for 
this project are illustrated in Figure~1.  Each involves alignment 
through the pulsar and another star, and they are the same as some 
of the slit orientations used by MacAlpine et al. (1996), in which extensive
N-rich and N-poor gas components were first identified.  
However, the spectral coverage extends further to the red for the 
optical observations presented here.

Near-infrared spectral coverage (from 7270 to 10030~\AA) was 
obtained during the nights of 2006 January 4 and 5, at the 2.7-m 
Harlan J. Smith telescope of the McDonald Observatory.  The Large 
Cassegrain Spectrograph was used with the CC1~1024$\times$1024 CCD, 
grating No. 42, and an RG610 blocking filter.  The dispersion 
through a 2\arcsec-wide slit was about 2.7~\AA~pixel$^{-1}$.  
The approximately 2\farcm6-long slit was placed in the positions shown 
with darker outlines in Figure~1 (with overlapping coverage 
to increase the length in the roughly north-south direction).

Observing conditions on all nights listed above were clear, 
and exposure times were at least 1~hour at every slit position.  
All of the optical and near-infrared two-dimensional images were carefully 
aligned along pixel columns or rows, and the data were reduced 
to relative flux against linear wavelength using IRAF\footnote{The Image Reduction and Analysis
Facility (IRAF) is distributed by the Association of Universities for
Research in Astronomy, Inc., under contract to the National Science Foundation.} software, 
along with observations of both lamp and moonlit-sky continua, wavelength calibration lamps, and 
spectrophotometric standard stars.  Sky spectral observations near the 
nebula were employed for removing night sky emission from the 
nebular spectra.

The goal was to identify and extract useful one-dimensional 
spectra at many positions spatially along the slits.  Resulting 
measured line intensities were corrected for differential dust extinction of E(B-V)~=~0.47 
(see Davidson \& Fesen 1985 and references therein) using the average 
interstellar extinction table from Osterbrock (1989).  

To obtain one-dimensional optical spectra, the dispersed radiation 
for every spatial pixel along the slits was carefully examined 
and compared with the associated two-dimensional image.  
Emission knots were identified, and appropriate combinations 
of two or three spatial pixels were averaged and extracted for 
our measurements.  In all cases, the individual spectral fluxes 
for averaged spatial pixels were required to be comparable 
(within 25\% of each other), and no single spatial pixels were 
used (in order to minimize the possibility of compromising the 
data by imperfect dispersion pixel alignment).  Whereas 
preliminary measurements were made by more than one person, 
all of the data employed here were ultimately, consistently 
measured by the first author.

A sample optical spectrum is presented in Figure~2.  This location was 
selected for illustration because it has relatively strong 
$[$O~I$]$, $[$Ar~III$]$, and $[$Ni~II$]$ lines.  Two emission 
systems (e.g., from filaments at the front and back of the 
expanding nebula) are often represented in the spectra; 
and here a lower-intensity, near-side system with very weak $[$N~II$]$ emission
can also be seen. Sometimes there is 
line blending, particularly in the wavelength range with 
$[$N~II$]$~$\lambda$6548, H$\alpha$, and $[$N~II$]$~$\lambda$6583.   
For such cases, 
IRAF deblending routines were employed, and occasionally 
the knowledge that $[$N~II$]$~$\lambda$6583~$\approx$~3~$[$N~II$]$~$\lambda$6548 was used 
in estimating line fluxes.  In general, repeated 
measurements of line fluxes were within 5\% of each other.  
Continuum placement in the optical spectra was reasonably 
straightforward, and the principal source of error was line 
blending.  Experiments showed that using 
various reasonable combinations of two or three spatial pixels  
at a filament location could lead to measurement 
differences up to 25\%, but these changes were always 
aligned along, or consistent with, the trends to be 
illustrated and discussed in \S~3; so specific pixel selection (following established guidelines) 
should not alter the conclusions.  Actual measurement errors for $\lambda$~$<$~7500~\AA\ lines
are estimated to be less than $\pm$10\%.

A sample near-infrared spectrum, illustrating strong $[$Ni~II$]$, $[$S~III$]$, 
and $[$C~I$]$ emission, is presented in Figure~3. Whereas there 
could have been more than one dynamically different emission system 
represented, line blending was never as much of a problem 
as in the H$\alpha$ and $[$N~II$]$ region.  Also, we note that these 
spectra generally involved larger numbers of averaged pixels 
than was the case for the optical spectra. Although the sky 
was clear, there were atmospheric seeing problems when the 
near-infrared data were obtained.  This and the larger slit 
width resulted in somewhat less resolution.  
In order to obtain broad wavelength coverage, 
potentially useful line-emission positions in each 
two-dimensional, near-infrared spectral image were 
identified and subsequently located as accurately as possible in the 
corresponding optical two-dimensional image.  Then 
optimal numbers of pixels were averaged to obtain 
consistent (for all wavelengths) one-dimensional spectra for these filaments.  
Because of the different slit widths and the fact that 
we could not expect exact correspondence between optical 
and near-infrared pixel groupings, it was necessary to 
normalize the spectra using the overlapped $[$Ni~II$]$~$\lambda$7378 
line in both wavelength ranges.  Although these normalization 
corrections could be quite accurate and were always less 
than a factor of 2, they still provided a significant 
potential source of error (estimated as much as 25\%) 
when comparing near-infrared and optical line intensities.

Another source of error, for one near-infrared 
line, is telluric water vapor absorption.  A plot of measured 
and reddening corrected values for $[$S~III$]$~$\lambda$9069 against
$[$S~III$]$~$\lambda$9531 shows a tightly defined linear relation 
(with a linear correlation coefficient greater than 0.99), 
as it should for emission from two transitions that arise 
from the same upper atomic level.   However, the predicted 
slope is about 2.6 for $[$S~III$]$~$\lambda$9531 on the vertical axis, according 
to the relevant transition probabilities given in the current 
NIST Atomic Spectra Database, whereas the slope of the line 
in our data is close to 2.0.   This type of situation has 
been reported before (Vermeij et al. 2002) and is caused 
by water vapor absorption in the 9531~\AA\ wavelength region.  
The absorption is clearly seen in our standard star spectra, 
and it should not significantly affect either the $[$S~III$]$~$\lambda$9069 
or $[$C~I$]$~$\lambda$9850 line measurements.  Because of this, in \S~3 
the well-measured $[$S~III$]$~$\lambda$9069 line will be used for examining 
spectral trends, rather than the stronger $[$S~III$]$~$\lambda$9531.

\section{TRENDS IN THE DATA}

All measured He~I~$\lambda$5876, $[$O~I$]$~$\lambda$6300, $[$N~II$]$~$\lambda$6583, 
$[$S~II$]$~$\lambda$6731, $[$Ar~III$]$~$\lambda$7136, $[$Ni~II$]$~$\lambda$7378, 
$[$S~III$]$~$\lambda$9069, 
and $[$C~I$]$~$\lambda$9850 emission line intensities were reddening 
corrected and normalized to the H$\alpha$ line.  
Then they were plotted against each other in various 
combinations, as we looked for trends that might provide 
new insights for understanding the physical conditions and 
chemical abundances in the Crab Nebula, with the future plan 
of developing improved photoionization models for the 
emitting gas.

\subsection{The Range of Nuclear Processing 
and Confirmation of Regions with Enhanced Products 
of Oxygen Burning}

As mentioned in the Introduction, MacAlpine et al. (1996)
identified regions in the Crab Nebula with either very 
strong or very weak $[$N~II$]$ emission (see their Figure~1).  
These gas regimes appear to be distinct both spatially 
and dynamically.  The $[$N~II$]$-weak 
gas often (but not always, as discussed below) shows 
unusually strong $[$S~II$]$ emission, and it was suggested 
that the latter probably indicates areas which contain 
products of oxygen-burning.  This
would be consistent with some stellar models of Woosley \& Weaver (1986b, 1995).  
MacAlpine et al. then used photoionization model analyses 
to derive overabundances of nitrogen by factors of 3 to 7 
(compared with solar) in $[$N~II$]$-strong regions and 
overabundances of Si, S, and Ar (assuming solar nitrogen) 
by factors of 10 to 20 for some $[$N~II$]$-weak areas.

Enhanced $[$S~II$]$ emission could also result from 
low-ionization, warm H$^{+}$$\rightarrow$H$^{o}$
transition zones in the emitting gas (Henry \& MacAlpine 1982), 
wherein S$^{+}$ ions are effectively collisionally excited 
by thermal electrons.  Therefore, the hypothesis for oxygen-burning 
products should be further investigated and convincingly 
demonstrated before its acceptance.  There are ways for 
examining this issue with the current data; and the most 
straightforward involve correlations between nitrogen and 
sulfur emission, between sulfur emission from different 
ions, or between emission from different elements 
expected to be produced together by the oxygen-burning process.

Figure~4 shows $[$N~II$]$~$\lambda$6583 plotted against 
$[$S~II$]$~$\lambda$6731.  We note the large and comparable range 
in intensities on both axes.  The nitrogen emission can 
be quite strong with very weak sulfur intensities, representing 
gas which has progressed no further than the CNO-cycle.  
Similarly, the strongest sulfur emission correlates only with 
weak nitrogen, suggestive of advanced processing through 
oxygen-burning.  It may also be seen that weak nitrogen 
does not always correspond with strong sulfur emission,
implying intermediate regions where some helium-burning has taken 
place and nitrogen has been converted into neon 
(see Pequignot \& Dennefeld 1983; Nomoto 1985; Henry 1986).
Regarding the latter point, infrared neon lines have
been observed in the Crab Nebula by Temim et al.
(2006).  Figure~9 of Temim et al. shows particularly strong
$[$Ne~II$]$~12.8~\micron\ emission in an area roughly 15\arcsec\ SW of the pulsar.
One of the slits used in this study crosses over an emitting filament
at that location, and averages of our measurements there are
$[$N~II$]$~$\lambda$6583/H$\alpha$~=~0.43 and $[$S~II$]$~$\lambda$6731/H$\alpha$~=~1.5.
As may be deduced from our Figure~4, these are low $[$N~II$]$ and modest $[$S~II$]$ values
that could be expected for a region where helium-burning and nitrogen depletion (to neon) have occurred,
but significant oxygen-burning has not taken place.

The data of Figure~4 illustrate the {\it range} of nuclear processing,
but not necessarily the relative {\it amounts} of material for the various
nucleosynthesis stages.  Because of line blending and other factors,
not all emitting positions could be measured and represented.

Figure~5 is a plot of $[$S~III$]$~$\lambda$9069 versus 
$[$S~II$]$~$\lambda$6731, which argues that the latter line
does not arise predominantly in optically-thick
H$^{+}$$\rightarrow$H$^{o}$ transition regions.  
Even with potential problems involving 
normalization of the near-infrared and optical spectra, 
there is a reasonably strong
correlation, with a linear correlation coefficient of 
0.75 for 37 points.  Therefore, since these different 
ionization stages (only one of which might be produced 
in an extended low-ionization zone) increase together, 
it would appear that strong sulfur emission observed 
in the Crab Nebula is mainly a result of enhanced 
sulfur abundance.

In addition to sulfur, other primary products of oxygen-burning 
include silicon and argon. Silicon emission is 
not strong in the optical and near-infrared regions, 
but $[$Ar~III$]$~$\lambda$7136 and $\lambda$7751 (weaker) can be measured 
in our data.   Therefore, in Figure~6 we plotted 
the correlation between $[$Ar~III$]$~$\lambda$7136 and $[$S~II$]$~$\lambda$6731 
for 182 locations from the optical 
slit spectra.  In this case, there is no potential 
problem with spectral normalization, and the correlation 
is quite high (linear correlation coefficient of 0.91).  
This confirms that the argon and sulfur emission must represent
abundances with a related origin.

Evidence presented here supports the hypothesis  
(MacAlpine et al. 1996; Strolger \& MacAlpine 1996) 
that certain regions in the Crab Nebula are heavily 
enriched with products of oxygen-burning.  This has 
important implications for understanding stellar models, 
and it will be explicitly considered in our follow-up 
photoionization model analyses.  For instance, as
pointed out by Henry (1993), high silicon and sulfur 
abundances would cause infrared fine-structure lines 
such as $[$Si~II$]$~34.8~\micron\ and $[$S~III$]$~33.6~\micron\ 
to become more important coolants for the gas, thereby 
influencing the line spectra in the optical and near-infrared.  
For lower electron temperature, longer-wavelength collisionally-excited
lines such as $[$S~II$]$~$\lambda\lambda$6716,6731 and $[$N~II$]$~$\lambda\lambda$6548,6583
would be enhanced at the expense of shorter wavelength lines like $[$O~II$]$~$\lambda\lambda$3726,3729.
Finally, we note that the above infrared fine-structure $[$Si~II$]$ and $[$S~III$]$
lines were recently observed to have significant 
intensities at some locations in the Crab Nebula (Temim et al. 2006).

Products of oxygen-burning are not unique to the Crab Nebula. Cassiopeia A, another young, 
collapse-driven, somewhat more massive supernova remnant
also exhibits extensive regions with high concentrations of silicon-group elements like sulfur and argon,
in which it is recognized that oxygen-burning has taken place while silicon-burning has been incomplete
(Chevalier \& Kirshner 1979; Willingale et al. 2002).

\subsection{$[$Ni~II$]$~$\lambda$7378 and $[$Ar~III$]$~$\lambda$7136 Line Trends:
Possible Implications for the Origin of Nickel}

In order to investigate observed very strong nickel 
emission in the Crab Nebula, 
we plotted $[$Ni~II$]$ intensity against every 
other measured line, and we found a particularly interesting correlation
with argon.
In Figure~7, all optical data are presented 
for the $[$Ni~II$]$~$\lambda$7378 and $[$Ar~III$]$~$\lambda$7136 lines.  
Points below the diagonal show a 
linear trend (correlation coefficient of 0.86) involving more highly processed gas, 
whereas points above the diagonal show the strongest $[$Ni~II$]$ emission and 
tend to represent regions where less nucleosynthesis has occurred.  

To investigate or highlight these apparent correlations further, we considered only
data for the very lowest and highest nitrogen emission.  
Following guidelines similar to those used by MacAlpine et al. (1996),
we identified the subset of points with measured 
$[$N~II$]$~$\lambda$6583~$<$~0.55~H$\alpha$
and $[$N~II$]$~$\lambda$6583~$<$~$[$S~II$]$~$\lambda$6731 
as being extreme ``low-N'' locations where 
advanced nuclear processing 
has taken place.  Similarly those points with 
$[$N~II$]$~$\lambda$6583~$>$~3~H$\alpha$ and 
$[$N~II$]$~$\lambda$6583~$>$~$[$S~II$]$~$\lambda$6731 were 
selected as being ``high-N,'' where nucleosynthesis stopped
with the CNO-cycle.  Figure~8 illustrates 
how these data appear in the $[$Ni~II$]$~$\lambda$7378 versus 
$[$Ar~III$]$~$\lambda$7136 plane, where filled squares represent  
high-N and open squares (with a +) denote low-N. 
The separation of trends in 
this plot is remarkable, with the low-N (more highly processed) positions 
having a linear correlation coefficient of 0.94.  

Some points below the diagonal of Figure~7 do not appear in Figure~8,
either because they do not have measurable $[$N~II$]$ emission or because
it is somewhat higher than the imposed limit for inclusion in Figure~8.   
We also note that, whereas emission from 
several pixel groups along one extended filament might 
conceivably create an almost linear structure of 
points in a diagram, the measurements 
for Figures~7 and 8 come from widely separated 
locations in all of the slits.

The linear correlation between $[$Ni~II$]$ and 
$[$Ar~III$]$ emission from the most highly processed gas 
could result from a type of ``alpha-rich freezeout'' 
(see Woosley \& Weaver 1995; Jordan et al. 2003), whereby
silicon-group elements in core-collapse supernovae may be heated by a shock wave
and broken down into nucleons and alpha particles.
Then, as the gas cools, these particles can reassemble into various stable iron-peak nuclei.
Another possibly contributing process has been discussed
by Thielemann \& Arnett (1985), who wrote:
``During O-burning, temperatures favor the photodisintegration
of heavy nuclei (produced by the s-process) into Fe-peak nuclei.
This is seen for $^{60}$Ni and partially for $^{62}$Ni and $^{58}$Fe,
depending on the prior neutron excess $\eta$.'' 

The above explanations would not 
apply to the steeper correlations in Figures~7 and 8, where 
$[$Ni~II$]$~$\lambda$7378 can be strongest in less-processed gas.  
This may be an indication that some
iron-peak, neutron-rich nuclei were removed from the 
surface of the neutron star.  For 
pulsars in general, this possibility was 
investigated by Ruderman \& Sutherland (1975),  
who concluded that extremely strong surface magnetic 
fields would not permit 
the release of heavy ions for most pulsars.  However, they 
also stated that the Crab Nebula's neutron star may be an 
exception to this rule because of very high surface temperature.
In this regard, we note that an extremely high temperature region on this 
young neutron star may have been detected by Weisskopf et al. (2004).
Furthermore, Michel et al. (1991) mapped an extensive north-south 
relativistic wind for the nebula, 
in the directions of the highest apparent 
concentrations of nickel (MacAlpine et al 1989).   
For line-emitting knots immersed in this wind, the highest $[$Ni~II$]$ emission was 
measured on the side facing the pulsar in each 
case (MacAlpine et al. 1994; MacAlpine \& Lawrence 1994). 
As Freiburghaus et al. (1999) have noted, the origins
of neutron-rich heavy elements are not yet well understood.
If ions can leave the surface of the neutron star in the 
Crab Nebula, then young pulsars in general could represent
sources for some heavy nuclei.

\subsection{On The Origin of Strong $[$C~I$]$ Emission}

Henry et al. (1984) measured the strength of $[$C~I$]$~$\lambda$9850 
in some filaments to be at least several times stronger 
than predicted by their photoionization models, and 
they suggested that previously neglected collisional 
excitation by H$^{0}$ may need to be considered as a 
potentially important process for production of this 
emission.

If we can understand where and how the $[$C~I$]$ emission 
is produced, we may be able to use it for estimating 
the carbon abundance relative to other elements.  
This, in turn, could provide additional useful insights 
into the extent of nuclear processing at various locations.  
As noted by Nomoto (1985) and Henry (1986), the amounts 
of carbon and oxygen would be depressed somewhat by 
CNO~processing and then would increase (above solar) 
as a result of helium-burning.  Also, Nomoto pointed 
out that an improved understanding of the carbon 
abundance, and therefore the amount of processing in 
the gas, can lead to a more accurate estimate of the 
stellar precursor mass.  Whereas carbon abundances 
derived from ultraviolet lines like C~IV~$\lambda$1549 for the 
Crab Nebula may be significantly influenced by the
way in which absorption of He~II~$\lambda$304 photons is 
considered (Eastman et al. 1985), that complication should not be 
an issue for the $[$C~I$]$~$\lambda$9850 line, since 
it arises away from the He~II-emitting gas.

New insight regarding the production of $[$C~I$]$ emission 
may be gained from examining its correlation with $[$O~I$]$.  
As previously shown in Figure~2, the Crab Nebula contains 
some locations with particularly strong $[$O~I$]$~$\lambda$6300 
emission, probably indicating the existence of broad 
H$^{+}$$\rightarrow$H$^{o}$ transition zones in the gas.  
The $[$O~I$]$ emission is known to be strengthened in these warm 
low-ionization regions (which could be expected for 
photoionization by a relatively flat synchrotron spectrum) 
because O$^{0}$ follows H$^{0}$ due to very effective charge 
exchange interactions.

Figure~9 illustrates the correlation between our measured 
$[$C~I$]$~$\lambda$9850 and $[$O~I$]$~$\lambda$6300 intensities.  
Although the plot is affected by the spectrum normalization 
procedure for near-infrared and optical wavelengths, the 
linear correlation coefficient is a significantly high 0.81.  
Therefore, we conclude that the strong $[$C~I$]$ emission is probably enhanced 
in extended ionization transition zones, by electron collisional excitation
and perhaps also by collisions involving H$^{0}$.  However, since 
the C and O contents of the gas could also have been depleted 
or increased {\it together} by the CNO-cycle or helium-burning, 
additional information is needed.

{\it Indirect} support for the idea that $[$C~I$]$ emission must be strengthened in 
low-ionization regions also comes from Figure~10, which shows $[$C~I$]$~$\lambda$9850 
plotted against $[$N~II$]$~$\lambda$6583.  There is a rather large 
(though not extreme) range of $[$N~II$]$ emission represented, 
and (except for the one very high point) $[$C~I$]$ looks 
random over an order of magnitude in the $[$N~II$]$ intensities.  
Since the latter may provide a rough representation of the 
amount of nucleosynthesis that has taken place, as 
discussed previously, it would appear that the strong measured 
$[$C~I$]$ emission is not directly correlated with nuclear processing.  
Clearly the roles of both abundance and ionization structure will be important 
avenues for exploration in further refinements of 
photoionization model computations.

\subsection{Exceptionally Strong He~I Emission}

The final abundance or line-intensity anomaly to be 
considered here involves the exceptionally high helium content 
derived by MacAlpine et al. (1989) for an apparent 
helium torus around the pulsar (see also Lawrence et al. 1995).  
Having measured dereddened He~I~$\lambda$5876~$\gtrsim$~H$\beta$, 
the results of photoionization analyses by Henry and MacAlpine (1982)
were used to infer a helium mass fraction around 95\% in this region.  

Henry \& MacAlpine (1982) also considered the possibility that 
He~I~$\lambda$5876 recombination lines could be significantly enhanced 
relative to hydrogen emission in low-ionization zones.
Because the photoionization cross 
section for He$^{0}$ scales roughly with frequency as $\nu$$^{-2}$,  
whereas that for H$^{0}$ scales more like $\nu$$^{-3}$, a high 
energy synchrotron radiation field can continue to ionize helium 
significantly beyond where hydrogen starts 
becoming neutral, thereby producing excess He~I recombination emission. 
This process could be especially 
important if the ionizing radiation flux is low (see, e.g., Shields 1974).
However, Henry \& MacAlpine found that photoionization 
models with relatively high values for the ionizing 
flux did a much better job of matching the majority of observed 
line intensities in the Crab Nebula, so they favored
very high derived abundance as the most plausible 
explanation for the strongest helium  
intensities. Now we have the opportunity to investigate this further.

As with $[$C~I$]$ emission, we examined 
the relation between He~I~$\lambda$5876 and $[$O~I$]$~$\lambda$6300 for 
our data, as shown in Figure~11.  
It may be seen that evidence for a meaningful correlation is lacking.  
Strong He~I emission is similarly represented at both
high and low $[$O~I$]$ intensities, so it is not 
enhanced primarily in low-ionization gas.  Furthermore, 
we considered the He~I~$\lambda$5876 correlation with $[$N~II$]$~$\lambda$6583 
in Figure~12.  Again, the strongest He~I emission is evident with both 
high and low $[$N~II$]$ measurements, so it apparently is not
directly tied to the overall progress of nuclear processing.
We conclude that the most reasonable explanation for 
the strongest He~I lines in the apparent torus around the pulsar 
is anomalously high abundance, the source of which is still 
in need of a plausible explanation.

\section{SUMMARY}
                 
We have measured emission of He~I~$\lambda$5876, $[$O~I$]$~$\lambda$6300, 
$[$N~II$]$~$\lambda$6583, $[$S~II$]$~$\lambda$6731, 
$[$Ar~III$]$~$\lambda$7136, $[$Ni~II$]$~$\lambda$7378, 
$[$S~III$]$~$\lambda$9069, $[$S~III$]$~$\lambda$9531, 
and $[$C~I$]$~$\lambda$9850 at many locations 
within the Crab Nebula.  The different line intensities 
(or subsets thereof) were plotted against each other 
in efforts to investigate correlations and improve our understanding of the range 
of nuclear processing in the gas, as well as the causes 
of exceptionally strong emission from $[$Ni~II$]$, $[$C~I$]$, 
and He~I.  We identified gas where nucleosynthesis
has not progressed significantly beyond the 
CNO-cycle, gas in which some helium-burning and nitrogen depletion have taken 
place, and regions where oxygen-burning has occurred.  
The anomalously strong observed $[$Ni~II$]$ emission may have 
two sources, in one case resulting from high temperature
and subsequent cooling in gas enriched with products of oxygen-burning, 
while in the other case possibly representing the release 
of nuclei from the neutron star surface.  Line correlations 
indicate that very strong $[$C~I$]$ emission
arises in low-ionization H$^{+}$$\rightarrow$H$^{o}$ transition regions.  
On the other hand, exceptionally strong He~I~$\lambda$5876 
does not show similar evidence of a low-ionization zone origin; and it does not appear to 
correlate with different levels of nuclear processing 
as represented by $[$N~II$]$ emission.  Therefore, the apparent
high-helium torus around the pulsar may be a distinct component
of the nebula.

\acknowledgments

We are grateful for generous financial and logistical support 
from Trinity University and the endowed Charles A. Zilker Chair position.
We also thank the staffs of the Michigan-Dartmouth-MIT Observatory
and the McDonald Observatory for providing excellent technical
assistance with this research.

\clearpage

\begin{figure}
\epsscale{1.0}
\plotone{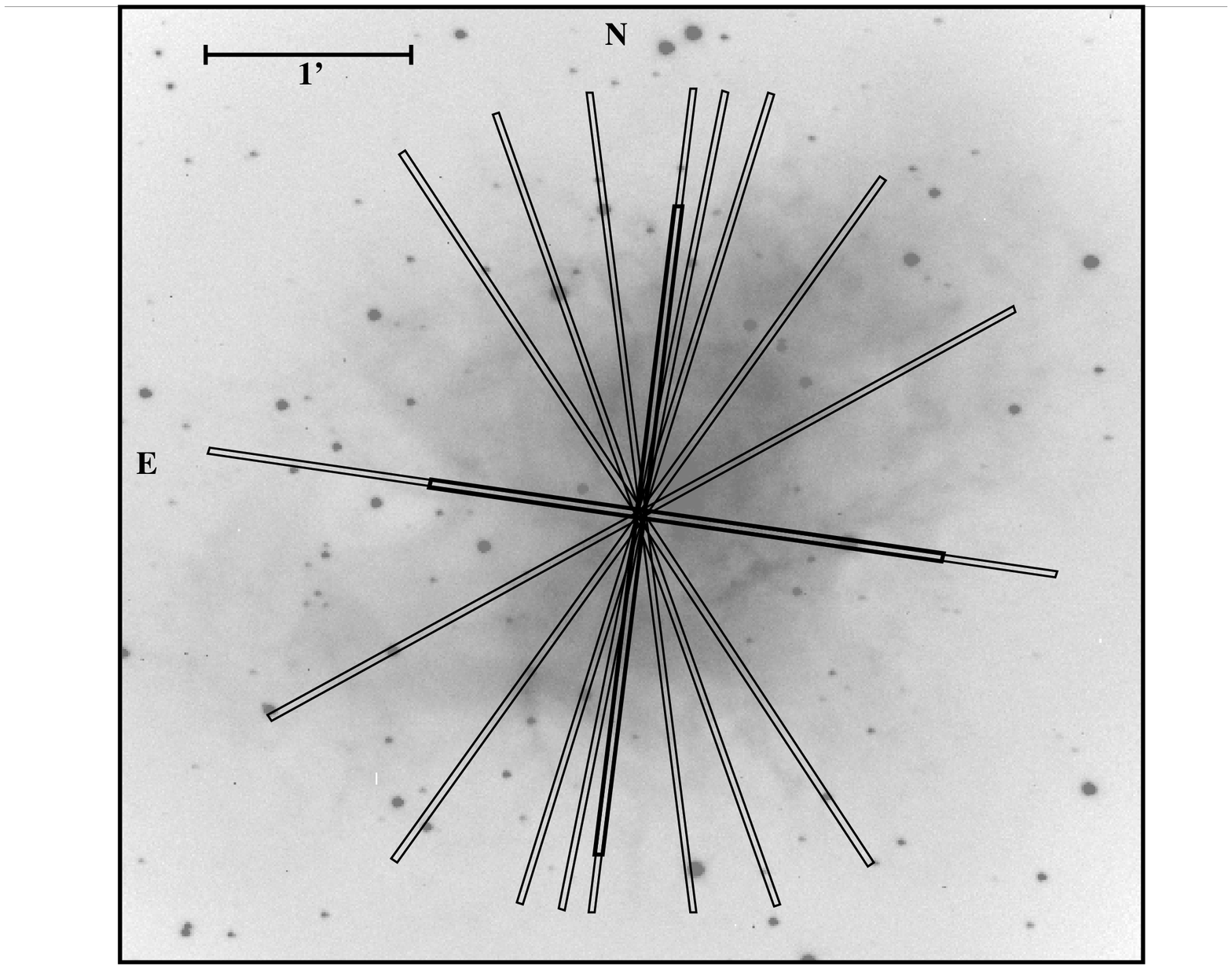}
\caption{Spectroscopic slit locations for this study,
superimposed on a V-band image of the Crab Nebula.
Each slit crosses over the pulsar and also goes through
a separate alignment star.  The nearly orthogonal, thicker,
shorter slit markers represent locations where near-infrared 
spectra were obtained. \label{fig1}}
\end{figure}
\clearpage

\begin{figure}
\epsscale{1.0}
\plotone{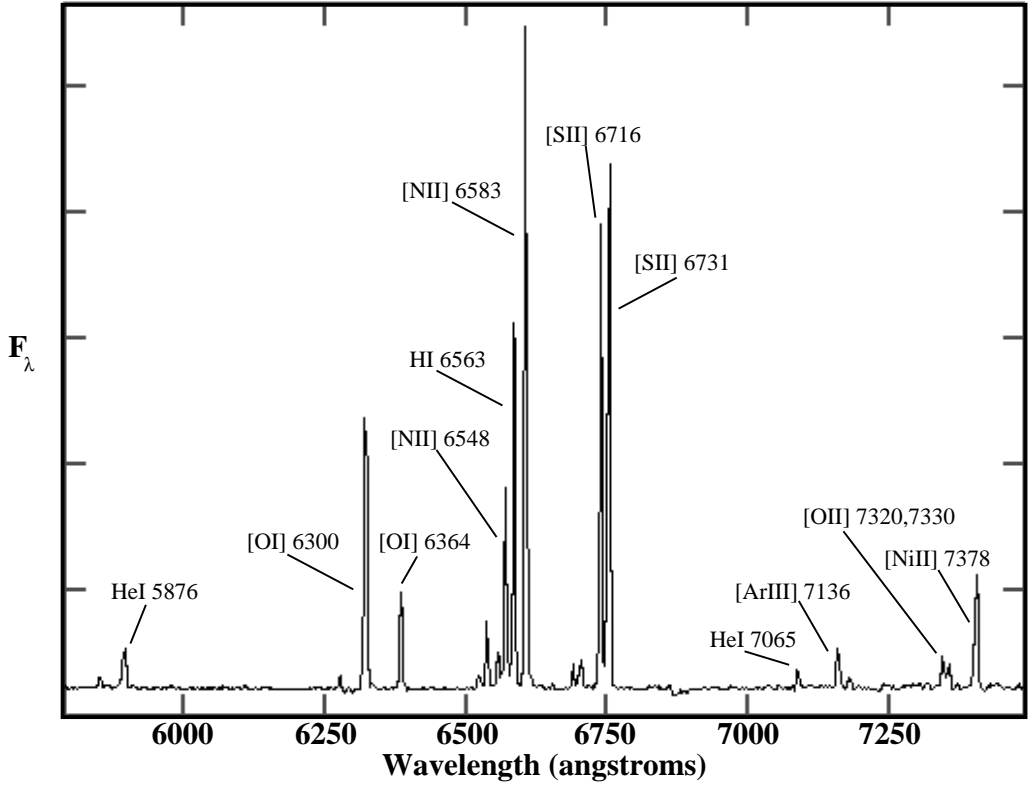}
\caption{A sample optical spectrum, with emission lines identified for gas with v~$\approx$~+900~km~s$^{-1}$.
Fainter lines for v~$\approx$~-1200~km~s$^{-1}$ may also be seen. 
The vertical axis is relative flux density. \label{fig2}}
\end{figure}
\clearpage

\begin{figure}
\epsscale{1.0}
\plotone{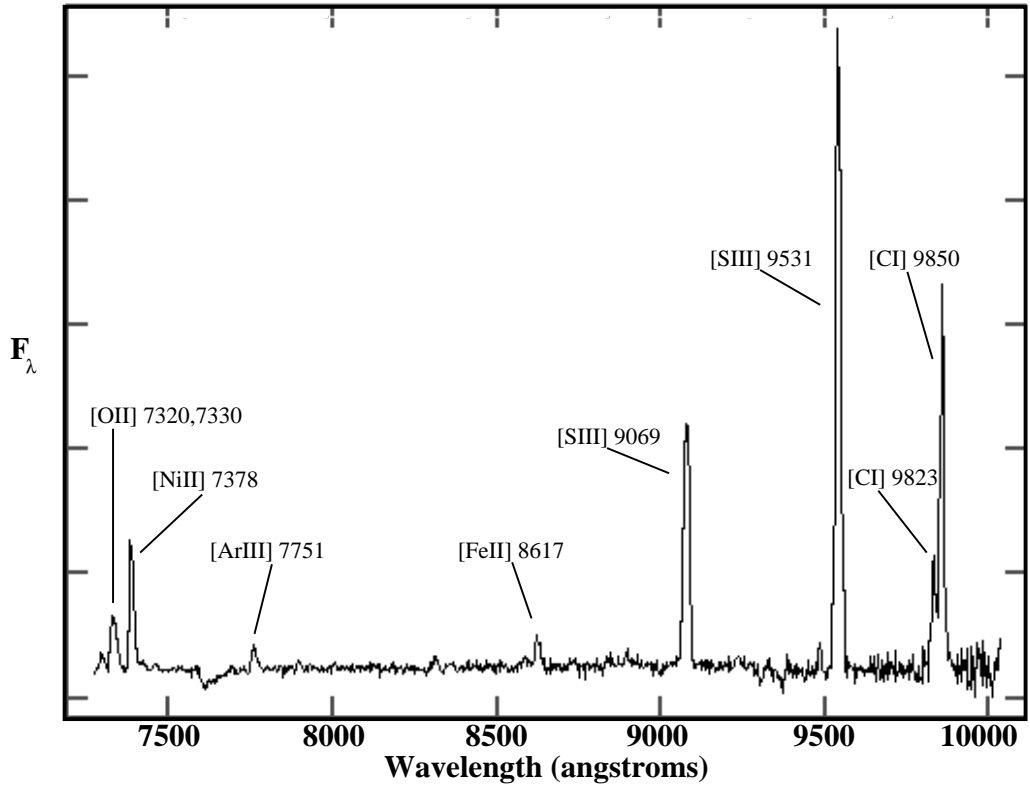}
\caption{A sample near-infrared spectrum with stronger lines indicated.
The vertical axis is relative flux density. \label{fig3}}
\end{figure}
\clearpage

\begin{figure}
\epsscale{1.0}
\plotone{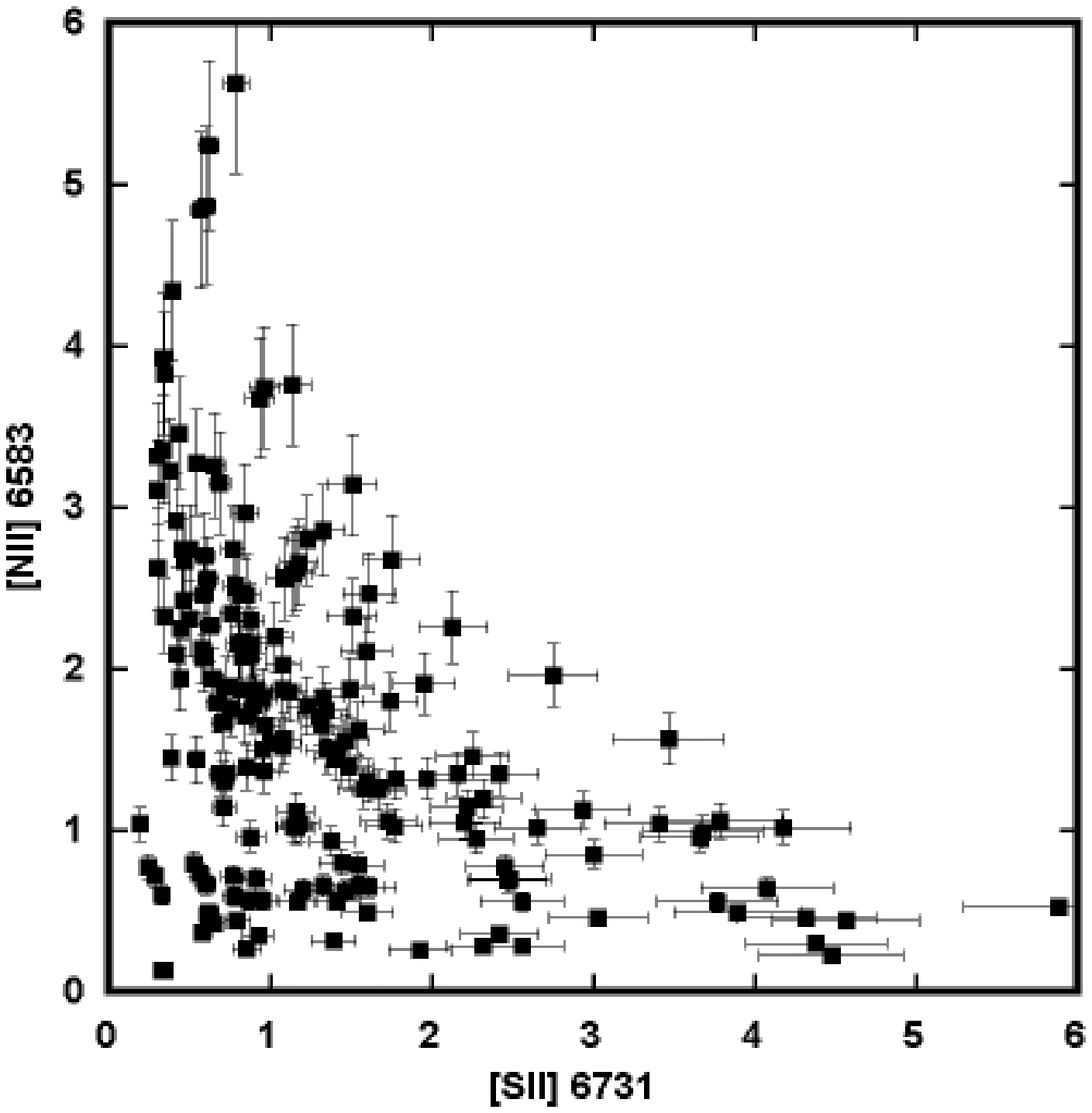}
\caption{Correlation between H$\alpha$-normalized
$[$N~II$]$~$\lambda$6583 and $[$S~II$]$~$\lambda$6731 
line intensities.  The error bars illustrate $\pm$~10\% uncertainties for both axes. \label{fig4}}
\end{figure}
\clearpage

\begin{figure}
\epsscale{1.0}
\plotone{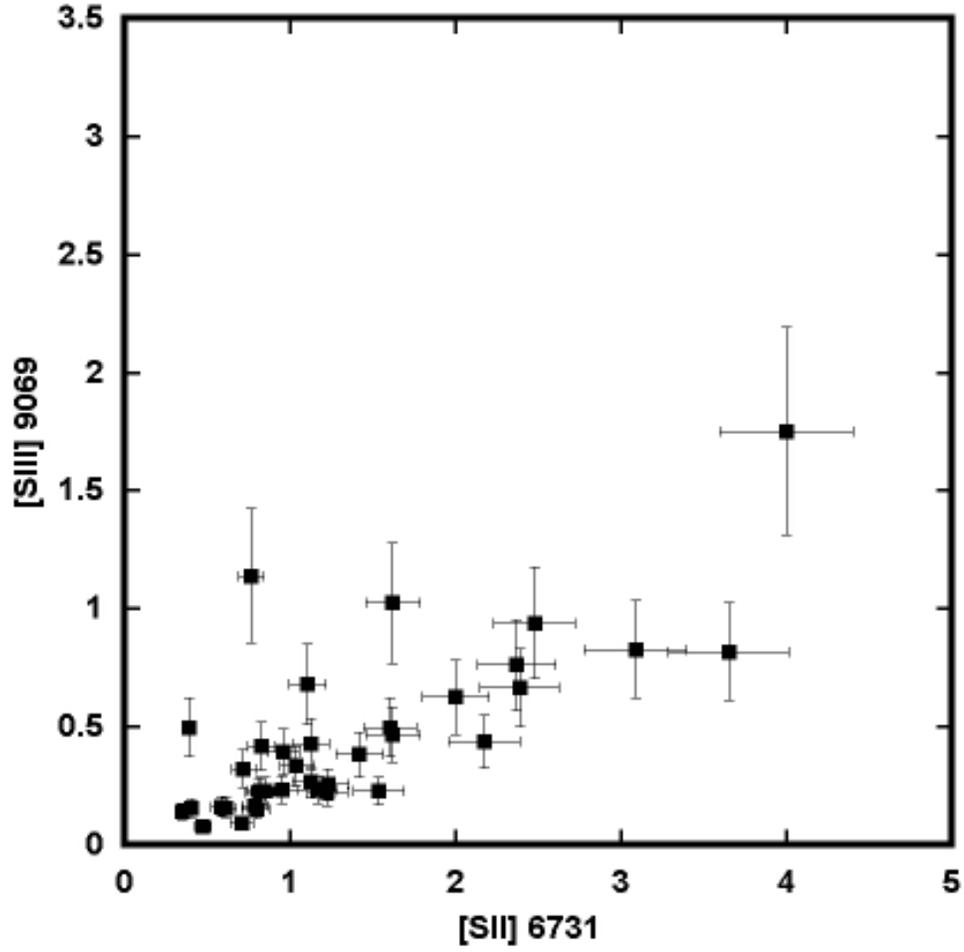}
\caption{Correlation between H$\alpha$-normalized
$[$S~III$]$~$\lambda$9069 and $[$S~II$]$~$\lambda$6731 
line intensities for the subset of combined near-infrared and optical measurements.
The error bars illustrate uncertainties of $\pm$~10\% for $[$S~II$]$~$\lambda$6731
and $\pm$~25\% for $[$S~III$]$~$\lambda$9069. 
 \label{fig5}}
\end{figure}
\clearpage

\begin{figure}
\epsscale{1.0}
\plotone{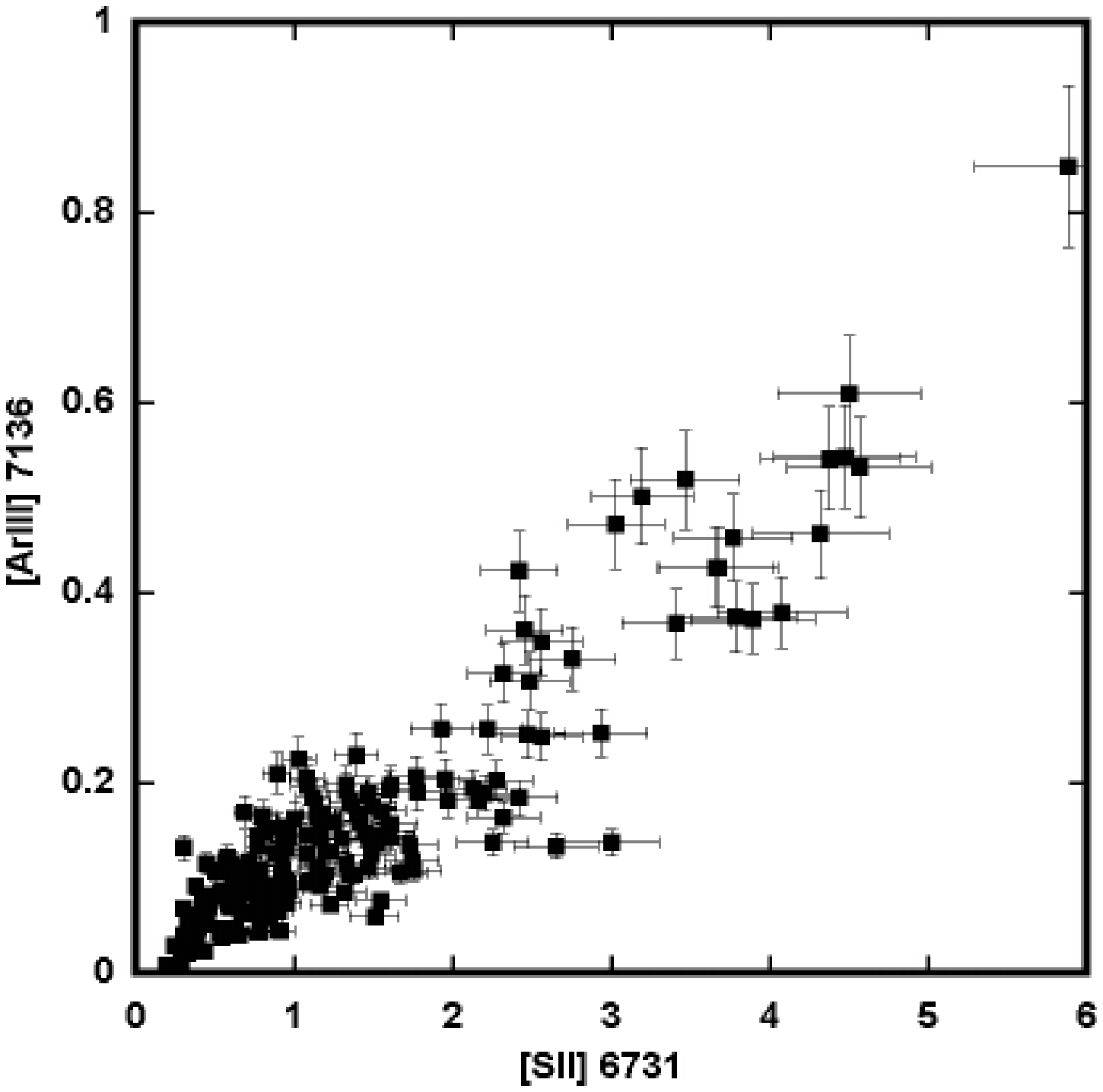}
\caption{Correlation between H$\alpha$-normalized $[$Ar~III$]$~$\lambda$7136 and $[$S~II$]$~$\lambda$6731
line intensities.  The error bars illustrate $\pm$~10\% uncertainties for both axes. 
\label{fig6}}
\end{figure}
\clearpage

\begin{figure}
\epsscale{1.0}
\plotone{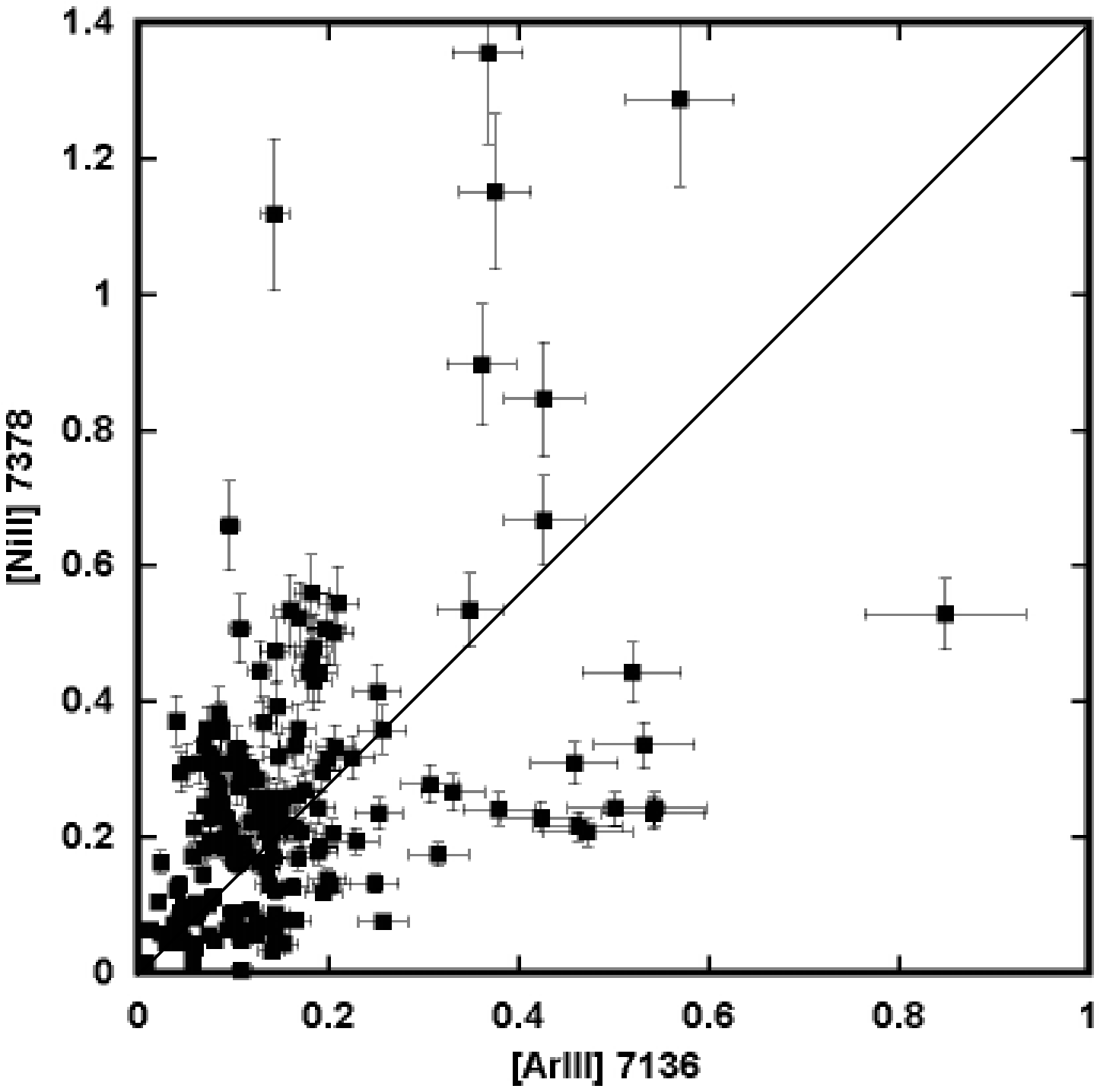}
\caption{Correlation between H$\alpha$-normalized $[$Ni~II$]$~$\lambda$7378 and $[$Ar~III$]$~$\lambda$7136 
line intensities. The error bars illustrate $\pm$~10\% uncertainties for both axes. 
\label{fig7}}
\end{figure}
\clearpage

\begin{figure}
\epsscale{1.0}
\plotone{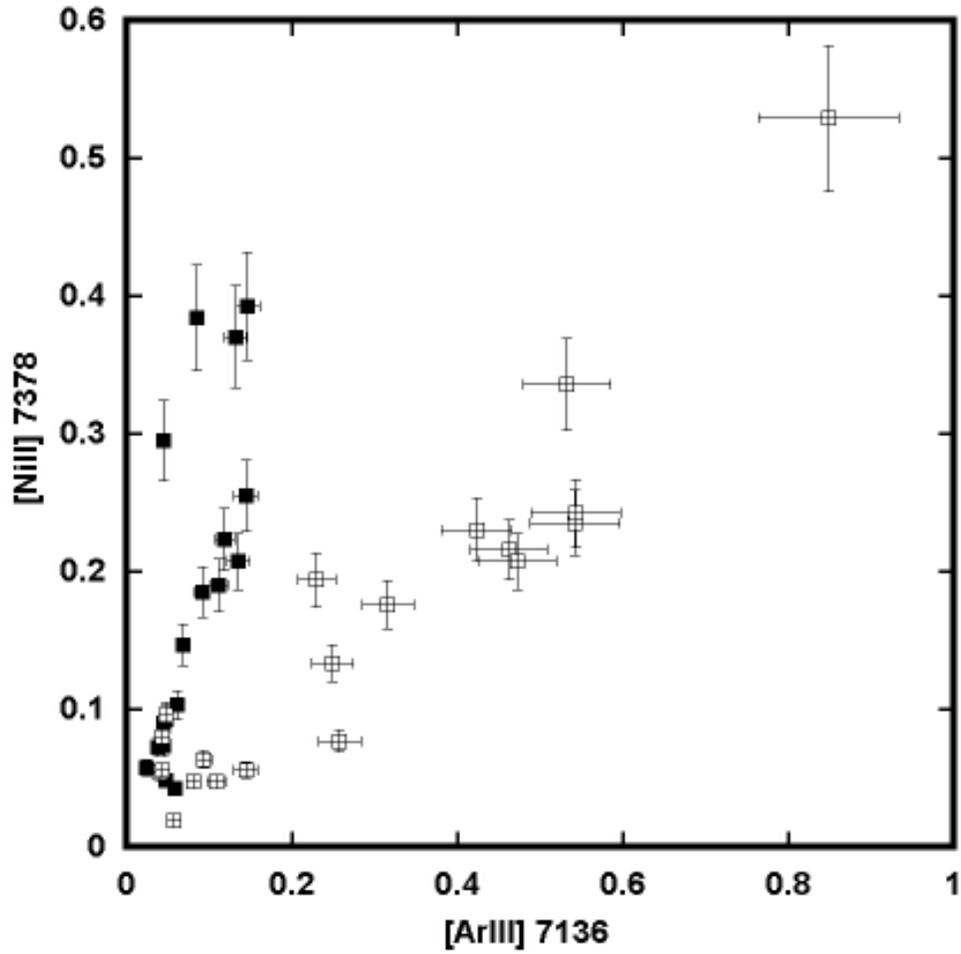}
\caption{Correlation between H$\alpha$-normalized $[$Ni~II$]$~$\lambda$7378 and $[$Ar~III$]$~$\lambda$7136 
line intensities for the subsets of extreme ``high-N'' (solid squares) and
``low-N'' (open squares with +) spectral data, as defined in the text. 
The error bars illustrate $\pm$~10\% uncertainties for both axes. 
\label{fig8}}
\end{figure}
\clearpage

\begin{figure}
\epsscale{1.0}
\plotone{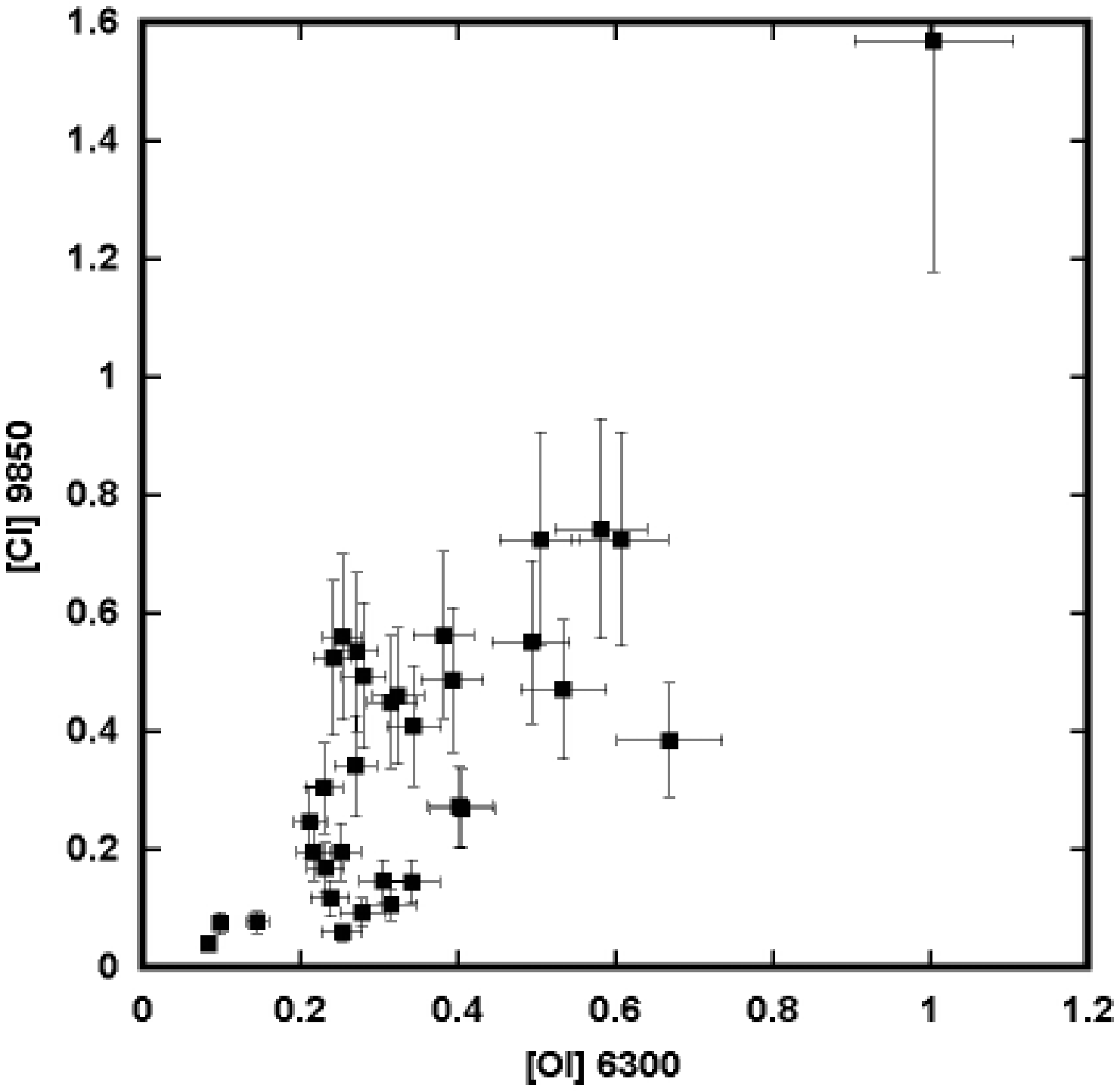}
\caption{Correlation between H$\alpha$-normalized $[$C~I$]$~$\lambda$9850 and $[$O~I$]$~$\lambda$6300 
line intensities for the subset of combined near-infrared and optical measurements. 
The error bars illustrate uncertainties of $\pm$~10\% for $[$O~I$]$~$\lambda$6300
and $\pm$~25\% for $[$C~I$]$~$\lambda$9850. 
\label{fig10}}
\end{figure}
\clearpage

\begin{figure}
\epsscale{1.0}
\plotone{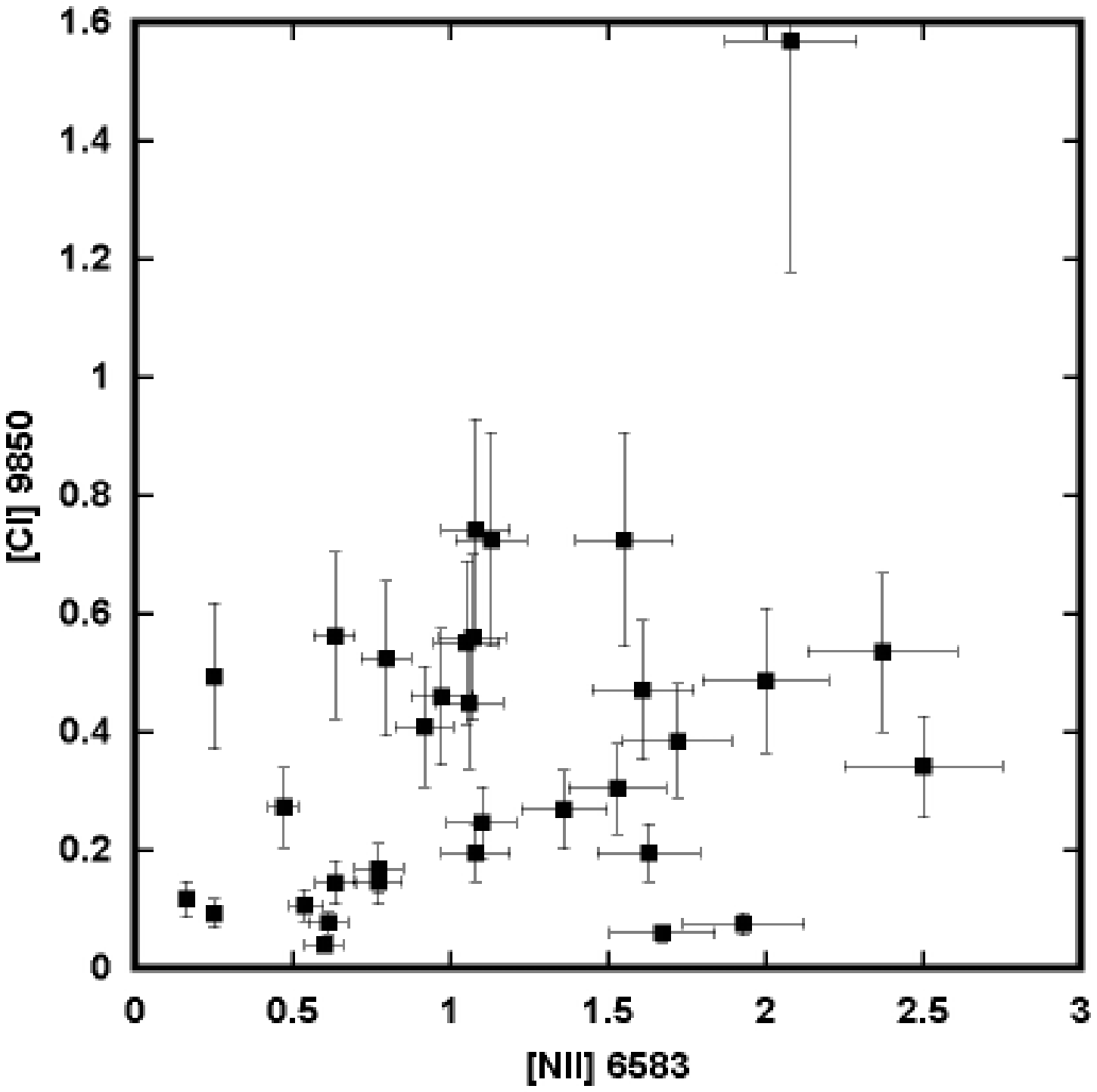}
\caption{Correlation between H$\alpha$-normalized $[$C~I$]$~$\lambda$9850 and $[$N~II$]$~$\lambda$6583 
line intensities for the subset of combined near-infrared and optical measurements. 
The error bars illustrate uncertainties of $\pm$~10\% for $[$N~II$]$~$\lambda$6583
and $\pm$~25\% for $[$C~I$]$~$\lambda$9850. 
\label{fig11}}
\end{figure}
\clearpage

\begin{figure}
\epsscale{1.0}
\plotone{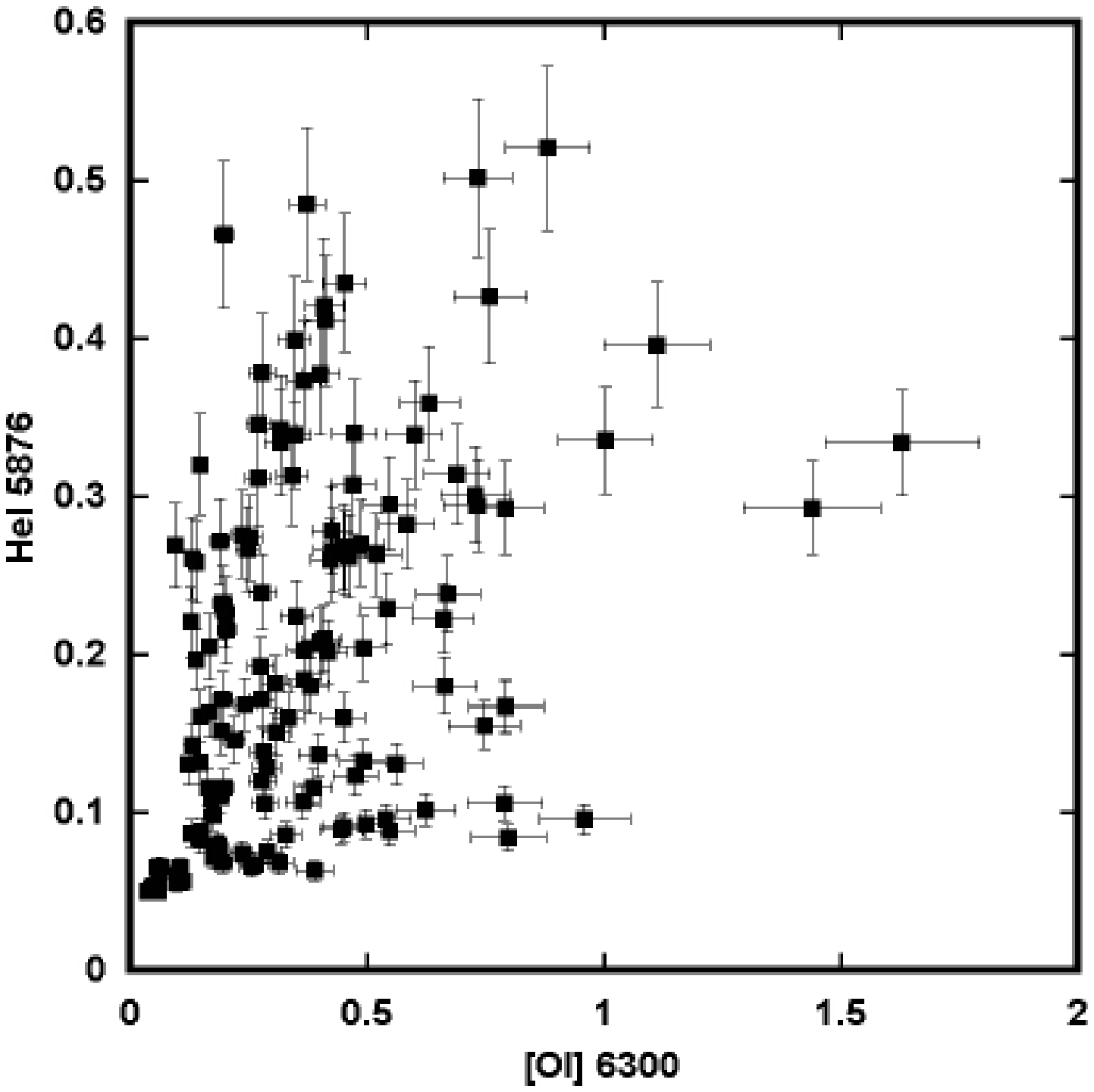}
\caption{Correlation between H$\alpha$-normalized He~I~$\lambda$5876 and $[$O~I$]$~$\lambda$6300
line intensities. 
The error bars illustrate $\pm$~10\% uncertainties for both axes. 
\label{fig12}}
\end{figure}
\clearpage

\begin{figure}
\epsscale{1.0}
\plotone{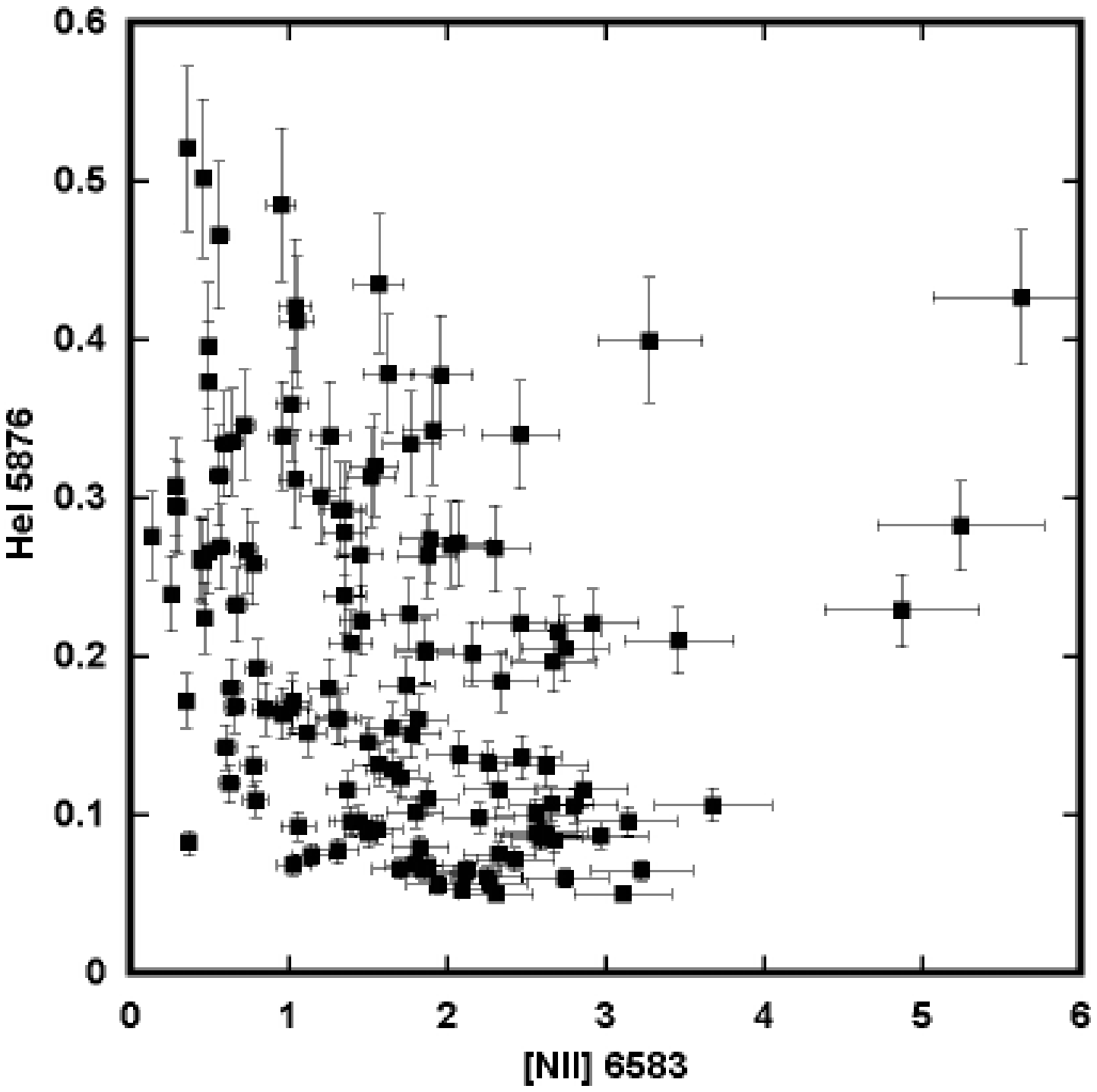}
\caption{Correlation between H$\alpha$-normalized He~I~$\lambda$5876 and $[$N~II$]$~$\lambda$6583 
line intensities. 
The error bars illustrate $\pm$~10\% uncertainties for both axes. 
\label{fig13}}
\end{figure}
\clearpage

\end{document}